\documentclass[fleqn,twoside]{article}

\usepackage{espcrc2}

\usepackage{epsfig}

\title{Wave functions and spectra from (S)DLCQ%
\thanks{Preprint no.\ UMN-D-01-9, to appear in the proceedings
of the International Workshop on Light-Cone Physics: Particles
and Strings, Trento, Italy, September 3-11, 2001.}%
}

\author{J.R. Hiller%
\address{Department of Physics \\
University of Minnesota-Duluth, Duluth, Minnesota 55812 USA}%
}
       
\begin{document}

\begin{abstract}
Applications of discretized light-cone quantization (DLCQ)
to (3+1)-dimensional Yukawa theory with Pauli--Villars
regulators and of supersymmetric DLCQ (SDLCQ) to (2+1)-dimensional
super Yang--Mills theory are discussed.  The ability of
these methods to provide wave functions as well as spectra
is emphasized.
\vspace{1pc}
\end{abstract}

\maketitle

\section{INTRODUCTION}

To be able to fully describe hadrons from first
principles in quantum chromodynamics (QCD), one needs 
to be able to compute wave functions and spectra.
Progress toward making such a computation in
the context of light-cone quantization has been
steady~\cite{DLCQreviews}.  Two particular approaches
will be discussed here.  Both make use of the discretized
light-cone quantization (DLCQ) technique pioneered
by Pauli and Brodsky~\cite{PauliBrodsky} for the
numerical solution of quantum field theories.
The approaches differ in how the theory is regulated;
one uses Pauli--Villars (PV) regularization~\cite{PauliVillars}
and the other supersymmetry.

The use of PV regularization with DLCQ in (3+1)-dimensional 
theories has been developed by Brodsky, Hiller, and 
McCartor~\cite{PV1,PV2,Yukawa}.  The initial work involved
simple many-body models~\cite{PV1,PV2}.  New work discussed
here and elsewhere~\cite{Yukawa} is for Yukawa theory
in a single-fermion truncation.  The essential idea is
to include PV particles in the DLCQ basis.  Cancellation
of ultraviolet infinities is then arranged by choosing
imaginary couplings or an indefinite metric.  In the
case of Yukawa theory there are choices to be made 
about the particle content of the PV sector.  The work
to be discussed here used three heavy scalars, two of which 
have negative norm.  An alternative now under investigation
is to use one heavy scalar and one heavy fermion, both with
negative norm.  This alternative was suggested by the work
of Paston {\em et al}.~\cite{Paston} and has the advantage
of being free of instantaneous fermion interactions.

Supersymmetric theories are simpler with respect to
regularization but require greater care in the numerical
discretization.  A key insight to the correct discretization
was made by Matsumura, Sakai, and Sakai~\cite{MatsumuraSakai}.
They noticed that by discretizing the supercharge $Q^-$
and constructing a discrete light-cone energy
$P^-$ from the superalgebra,
one could retain supersymmetry in the discrete approximation,
now called SDLCQ~\cite{SDLCQreview}.  Ordinary DLCQ discretizes
$P^-$ directly and recovers supersymmetry only in the
infinite resolution limit.  The SDLCQ technique has been
refined and applied by Pinsky and collaborators~\cite{SDLCQreview},
initially in 1+1 dimensions, but now also in 2+1 
dimensions~\cite{SYM2+1,SYMnew}.  Recent work on (2+1)-dimensional
supersymmetric Yang--Mills theory is discussed here and
elsewhere~\cite{SYMnew,ThisVolume}.

Both approaches rely on DLCQ.  All light-cone momentum
variables are discretized, with $p^+\rightarrow n\pi/L$
and $\vec{p}_\perp\rightarrow \vec{n}_\perp\pi/L_\perp$,
in terms of longitudinal and transverse length scales
$L$ and $L_\perp$.  The integrals over wave functions
that make up the mass eigenvalue problem $H_{\rm LC}\Phi=M^2\Phi$
are then approximated by the trapezoidal quadrature rule.  This
yields a matrix eigenvalue problem which is typically
quite large but also quite sparse.  Lanczos 
techniques~\cite{Lanczos} are used to extract 
eigenvalues and eigenvectors for the lowest states,
even in the case of an indefinite metric~\cite{Yukawa}.

Because the longitudinal momentum is always positive,
there exists a positive integer $K$, called the
(harmonic) resolution~\cite{PauliBrodsky}, such that
the total longitudinal momentum is $P^+=K\pi/L$ and
momentum fractions are given by $x=n/K$.  Wave functions
and the mass eigenvalue problem, where $H_{\rm LC}=P^+P^-$,
are naturally expressed in terms of momentum fractions
and the resolution $K$.  Hence $L$ disappears, and 
$K$ effectively takes its place as the resolution scale.
The transverse scale $L_\perp$ is set by a momentum cutoff
and a transverse resolution.

The remainder of this paper contains brief discussions
of recent work on Yukawa theory~\cite{Yukawa} in
Sec.~\ref{sec:Yukawa} and super Yang--Mills theory~\cite{SYMnew}
in Sec.~\ref{sec:SYM}.  A summary and discussion of future
work are given in Sec.~\ref{sec:Future}.

\section{YUKAWA THEORY}  \label{sec:Yukawa}

When terms involving antifermions are eliminated, the
Yukawa light-cone Hamiltonian becomes~\cite{McCartorRobertson}
\begin{eqnarray}
\lefteqn{H_{\rm LC}=
   \sum_{\underline{n},s}
      \frac{M^2+\delta M^2+(\vec{n}_\perp \pi/L_\perp)^2}{n/K}
          b_{\underline{n},s}^\dagger b_{\underline{n},s}}
\nonumber \\
   &&+\sum_{\underline{m}}
          \frac{\mu^2+(\vec{m}_\perp \pi/L_\perp)^2}{m/K}
              a_{\underline{m}}^\dagger a_{\underline{m}} 
\nonumber \\
   && +\frac{g\sqrt{\pi}}{2L_\perp^2}
          \sum_{\underline{n}\underline{m}s}\frac{1}{\sqrt{m}}
     \left[\frac{\vec{\epsilon}_{-2s}^{\,*}\cdot\vec{n}_\perp}{n/K}\right.
\nonumber \\
   &&  +\left.
    \frac{\vec{\epsilon}_{2s}\cdot(\vec{n}_\perp+\vec{m}_\perp)}{(n+m)/K}
                          \right]
     b_{\underline{n}+\underline{m},-s}^\dagger b_{\underline{n},s} 
                                            a_{\underline{m}} 
\nonumber \\
  && \rule{0.5in}{0mm} + \mbox{h.c.} 
\nonumber \\
   && +\frac{Mg}{\sqrt{8\pi}L_\perp}
             \sum_{\underline{n}\underline{m}s}\frac{1}{\sqrt{m}}
     \left[\frac{1}{n/K}\right.  \\
  &&  +\left.\frac{1}{(n+m)/K}\right]
     b_{\underline{n}+\underline{m},s}^\dagger b_{\underline{n},s} 
                a_{\underline{m}}    + \mbox{h.c.} 
\nonumber \\
   && +\frac{g^2}{8\pi L_\perp^2}\sum_{\underline{n}\underline{m}\underline{m}'s}
         \frac{1}{\sqrt{mm'}} 
\nonumber \\
&& \times
 \left[b_{\underline{n}+\underline{m}+\underline{m}',s}^\dagger 
         b_{\underline{n},s} a_{\underline{m}'} a_{\underline{m}}\frac{1}{(n+m)/K}
              + \mbox{h.c.}  \right. 
\nonumber \\
    && + b_{\underline{n}+\underline{m}-\underline{m}',s}^\dagger 
           b_{\underline{n},s} a_{\underline{m}'}^\dagger a_{\underline{m}} 
\nonumber \\
  &&    \left.\times\left(\frac{1}{(n-m')/K}+\frac{1}{(n+m)/K}\right)\right]
\nonumber 
\end{eqnarray}
where
\begin{eqnarray}
\left\{b_{\underline{n},s},b_{\underline{n}',s'}^\dagger\right\}
     &=&\delta_{\underline{n},\underline{n}'} \delta_{s,s'}\,, \\
\left[a_{\underline{m}},a_{\underline{m}'}^\dagger\right]
          &=&\delta_{\underline{m},\underline{m}'}\,,
\nonumber
\end{eqnarray}
and
\begin{equation}
\vec{\epsilon}_\lambda=-\frac{1}{\sqrt{2}}(\lambda,i)\,. 
\end{equation}
The one-loop contribution to the fermion self energy is~\cite{PV1}
\begin{eqnarray}
\lefteqn{I(\mu^2,M^2)\simeq
\frac{\pi}{\mu^2}\left[\left( \frac{\Lambda^2}{2} - \mu^2 \ln \Lambda^2
    \right.\right. }
\nonumber \\
&&   \rule{1in}{0mm}
 \left.  + \mu^2 \ln \mu^2 - \frac{\mu^4}{2\Lambda^2}\right)   \\
&& + M^2\left( 3 \ln\Lambda^2 - 3 \ln \mu^2 - \frac{9}{2}
                           + \frac{5\mu^2}{\Lambda^2}\right) 
\nonumber \\
&& +  \left.M^4\left(\frac{2}{\mu^2}\ln (M^2/\mu^2)
                 +\frac{1}{3\mu^2}-\frac{1}{2\Lambda^2}\right)\right]\,,
\nonumber 
\end{eqnarray}
with $\Lambda$ a cutoff.  For this to be finite and consistent with
zero in the $M\rightarrow 0$ chiral limit, we need to perform three
subtractions.  Therefore we add to the Hamiltonian three heavy PV 
scalars~\cite{ThreePV} with masses $\mu_i$ and couplings $\xi_i g$ 
determined by
\begin{eqnarray} 
\lefteqn{1+\sum_{i=1}^3 (-i)^i\xi_i=0\,, \;\;
\mu^2+\sum_{i=1}^3 (-i)^i\xi_i\mu_i^2=0\,,} \nonumber \\
&& \sum_{i=1}^3 (-i)^i\xi_i\mu_i^2\ln(\mu_i^2/\mu^2)=0\,.
\end{eqnarray}
These constraints guarantee the desired one-loop subtractions
if the norm of the i-th PV boson is $(-1)^i$.
Fermion self-induced inertia terms have not been included
because they cancel once the heavy scalars are added.

The state vector is
\begin{eqnarray}
\lefteqn{\Phi_\sigma=\sqrt{16\pi^3P^+}\sum_{n_0,n_1,n_2,n_3=0}^\infty
     \int\frac{dp^+d^2p_\perp}{\sqrt{16\pi^3p^+}}} \\
&& \times   \prod_{j=1}^{n_{\rm tot}}
     \int\frac{dq_j^+d^2q_{\perp j}}{\sqrt{16\pi^3q_j^+}} 
        \sum_s\delta(\underline{P}-\underline{p}
                     -\sum_{j}^{n_{\rm tot}}\underline{q}_j)
\nonumber \\
   &  & \times 
     \phi_{\sigma s}^{(n_i)}(\underline{q}_j;\underline{p})
     \frac{1}{\sqrt{\prod_i n_i!}}b_{\underline{p}s}^\dagger
      \prod_j^{n_{\rm tot}} a_{i_j\underline{q}_j}^\dagger |0\rangle \,,
\nonumber 
\end{eqnarray}
with $n_{\rm tot}=\sum_i n_i$ and normalization
\begin{equation}
\Phi_\sigma^{\prime\dagger}\cdot\Phi_\sigma
=16\pi^3P^+\delta(\underline{P}'-\underline{P})\,. 
\end{equation}

To cancel an infrared singularity in the instantaneous fermion term, 
we add an effective interaction that represents the contribution of
the missing Z graph.  The added term is built diagrammatically
from the pair production and annihilation terms
\begin{eqnarray}
\lefteqn{{\cal P}_{\rm pair}^-= \frac{g}{2L_\perp\sqrt{L}}
   \sum_{\underline{p}\underline{q}si}
  \left[\frac{\vec{\epsilon}_{-2s}\cdot\vec{p}_\perp}{p^+\sqrt{q^+}}
    \right.} \nonumber \\
 &&+\left.\frac{\vec{\epsilon}_{2s}^{\,*}\cdot(\vec{q}_\perp-\vec{p}_\perp)}
                    {(q^+-p^+)\sqrt{q^+}}\right]
     b_{\underline{p},s}^\dagger d_{\underline{q}-\underline{p},s}^\dagger 
\xi_i a_{i\underline{q}}
     +\mbox{h.c.} \nonumber \\
   && +\frac{Mg}{2L_\perp\sqrt{2L}}
   \sum_{\underline{p}\underline{q}si}
    \left[\frac{1}{p^+\sqrt{q^+}}  \right.  \\
&& -\left.\frac{1}{(q^+-p^+)\sqrt{q^+}}\right]
     b_{\underline{p},s}^\dagger d_{\underline{q}-\underline{p},-s}^\dagger 
\xi_i a_{i\underline{q}}
     +\mbox{h.c.}\,,  \nonumber
\end{eqnarray}
and the denominator for the intermediate state
\begin{eqnarray}
\frac{M^2}{P^+}&-&p_{\rm spectators}^- -\frac{M^2+p_\perp^{\prime 2}}{p'^+}\\
 & -&\frac{M^2+(\vec{q}_\perp^{\,\prime}-\vec{p}_\perp)^2}{q'^+-p^+}
          -\frac{M^2+p_\perp^2}{p^+}\,. \nonumber
\end{eqnarray}

The bare parameters of the Hamiltonian, $g$ and $\delta M^2$,
are determined by input of ``data.''
The mass $M$ of the dressed single-fermion state is held fixed.  This
is imposed by rearranging the mass eigenvalue problem
into an eigenvalue problem for $\delta M^2$:
\begin{eqnarray}
\lefteqn{x\left[M^2- \frac{M^2+p_\perp^2}{x}
   -\sum_j\frac{\mu_j^2+q_{\perp j}^2}{y_j}\right] \tilde{\phi}} \\
&&-\int\prod_j dy'_j d^2q'_{\perp j}\sqrt{xx'}{\cal K}\tilde{\phi}'
=\delta M^2\tilde{\phi}\,,  \nonumber
\end{eqnarray}
where ${\cal K}$ represents the original kernel and amplitudes are related 
by $\phi=\sqrt{x}\tilde{\phi}$.

To fix the coupling we use
$\langle :\!\!\phi^2(0)\!\!:\rangle
\equiv\Phi_\sigma^\dagger\!:\!\!\phi^2(0)\!\!:\!\Phi_\sigma$.
From a numerical solution it can be computed fairly efficiently in a sum
similar to the normalization sum
\begin{eqnarray}
\lefteqn{\langle :\!\!\phi^2(0)\!\!:\rangle
        =\sum_{n_i=0}^\infty \int \prod_j^{n_{\rm tot}}
                          \,dq_j^+d^2q_{\perp j} \sum_s (-1)^{(n_i)}}
    \\
  & & \times \left(\sum_{k=1}^n \frac{2}{q_k^+/P^+}\right)
       \left|\phi_{\sigma s}^{(n_i)}(\underline{q}_j;
       \underline{P}-\sum_j\underline{q}_j)\right|^2\,.
\nonumber
\end{eqnarray}
The constraint on $\langle :\!\!\phi^2(0)\!\!:\rangle$ is satisfied by 
solving it simultaneously with the eigenvalue problem.

\begin{figure}[htb]
\psfig{file=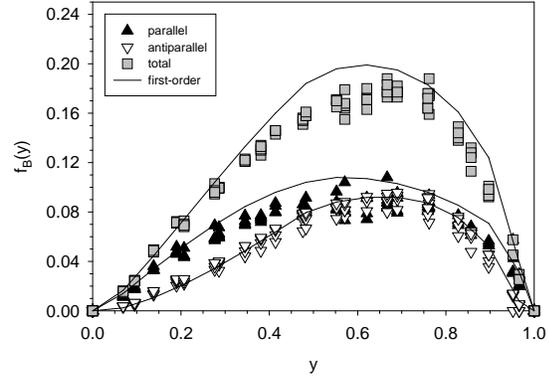,width=7.5cm}
\caption{\label{fig:StructFn} The boson structure function 
$f_B$ at various numerical resolutions for 
$\langle:\!\!\phi^2(0)\!\!:\rangle=0.5$, 
with $M=\mu$, cutoff $\Lambda^2=50\mu^2$, 
and Pauli--Villars masses $\mu_1^2=10\mu^2$, $\mu_2^2=20\mu^2$, 
and $\mu_3^2=30\mu^2$.  The solid line is from first-order perturbation
theory. } 
\end{figure}

With the parameters fixed, we can compute various quantities.
Those considered include structure functions, the
form factor slope at zero momentum transfer, average 
numbers of constituents, and average constituent momenta.
A representative plot of the bosonic structure function
\begin{eqnarray}
\lefteqn{f_B(y)\equiv\sum_{n_i=0}^\infty\sum_s
   \int\,\prod_jdq_j^+d^2q_{\perp j} (-1)^{(n_i)}\sum_{k=1}^{n_0}}
\nonumber  \\
   &&\times  \delta(y-q_k^+/P^+)
      \left|\phi_{\sigma s}^{(n_i)}(\underline{q}_j;
                      \underline{P}-\sum_i\underline{q}_j)\right|^2\,,
\end{eqnarray}
is given in Fig.~\ref{fig:StructFn}.
Additional results can be found in Ref.~\cite{Yukawa}.  For good
results at stronger couplings, where first-order perturbation
theory is insufficient, rather high resolution was required,
with $K=21$ to 39 and as many as 15 transverse momentum points.
This resolution was achieved by limiting the number of
constituents to 3, after verifying that the contribution 
from higher sectors was sufficiently small.

\section{SUPER YANG--MILLS THEORY} \label{sec:SYM}

The Lagrangian for supersymmetric SU(N) Yang--Mills
theory in 2+1 dimensions is
\begin{equation}
{\cal L}=\mbox{tr}(-\frac{1}{4}F^{\mu\nu}F_{\mu\nu}+
             {\rm i}{\bar\Psi}\gamma^\mu D_\mu\Psi)\,,
\end{equation}
where $F_{\mu\nu}=\partial_\mu A_\nu-\partial_\nu A_\mu
+ig[A_\mu,A_\nu]$ and $D_\mu\Psi=\partial_\mu\Psi+ig[A_\mu,\Psi]$.
We work in light-cone gauge ($A^+=0$) and the large-$N$ limit.
The dynamical fields are $\phi\equiv A_2$ and 
$\psi\equiv 2^{-1/4}(1+\gamma^5)\Psi$.  The supercharge
$Q^-$ is
\begin{eqnarray}
\lefteqn{Q^-=2^{3/4}\int dx^-
    \int_0^l dx_\perp\mbox{tr}\left[\partial_\perp\phi\psi \right.}
\nonumber \\
  &&+  \left.g_{\rm YM}\left({\rm i}[\phi,\partial_-\phi]
            +2\psi\psi\right)\frac{1}{\partial_-}\psi\right]\,.
\end{eqnarray}
and $P^-$ is given by $\{Q^-,Q^-\}=2\sqrt{2}P^-$.
By discretizing $Q^-$ and computing $P^-$ from this
anticommutator, supersymmetry is exactly preserved in
the numerical approximation~\cite{MatsumuraSakai}.  In
addition to supersymmetry, we have transverse parity and
the Kutasov $T$ symmetry~\cite{Kutasov}.  The mass
eigenvalue problem can then be solved separately in each
of the 8 symmetry sectors.  In the largest calculation
to date~\cite{SYMnew}, each sector contained roughly 
230,000 basis states.

A variety of results can be seen in Ref.~\cite{SYMnew}.
They include a number of studies of the coupling dependence
of the mass values, as well as outcomes for structure
functions.  One striking feature is that, except for
weak coupling, the average particle count for almost all
eigenstates is at or near the maximum allowed by the
resolution, even for the highest resolution considered.
This means that for strong coupling the method does not
capture all of the important pieces of the true eigenstate.
Such behavior is likely to be a consequence of dealing
with massless constituents without introducing a mass
scale through symmetry breaking or other means.

\begin{figure}[ht]
\psfig{file=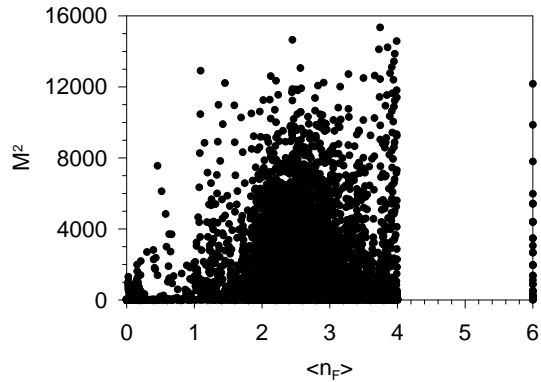,width=7.5cm}
\caption{Bound-state masses squared $M^2$ in units of $4 \pi^2 / L_\perp^2$
for bosonic states  
as functions of the average number of fermions $\langle n_F\rangle$ for
$K=6$.  Several different values of the coupling are included.
The symmetry sector is one with positive $T$ symmetry and
positive parity.
\label{fig:<nF>}}
\end{figure}

Another striking feature is that the average number of
fermions in bosonic states has been observed to have a 
gap between 4 and 6.  This could only be checked for
resolutions where a full diagonalization could be done.
However, as can be seen in Fig.~\ref{fig:<nF>},
the signal is quite clean for $K=6$ and 3 transverse
momentum modes, where the average number is never above
4 unless it is precisely 6, the maximum allowed; below 4
there are no gaps at all.

\section{FUTURE WORK}\label{sec:Future}

Both ordinary DLCQ with Pauli--Villars particles and SDLCQ
provide the means to compute masses and wave functions for 
eigenstates in multi-dimensional quantum field theories.
These methods will continue to be explored in various contexts,
leading eventually to consideration of (supersymmetric) QCD.
In fact, Paston {\em et al}.~\cite{PastonQCD} have already
obtained a PV-like regularization of QCD that could, in
principle, be solved by DLCQ; however, with present computing power
the number of fields is probably too large for meaningful
calculations.

With QCD as a goal, work on the PV approach will next turn 
to the alternative regularization of Yukawa theory, with
one heavy scalar and one heavy fermion~\cite{Paston},
in both the one and two-fermion Fock sectors.  Quantum electrodynamics
will also be considered, as a first application to a gauge
theory and as something of interest in its own right.

The next step for work with SDLCQ is to include a 
Chern--Simons term in super Yang--Mills theory and
thereby give each constituent a nonzero mass.  The
extension of the method to 3+1 dimensions is also
important, as is consideration of supersymmetry breaking.

\section*{ACKNOWLEDGMENTS}
\noindent
The work reported here was done in collaboration 
with S.J. Brodsky and G. McCartor
and with S.S. Pinsky and U. Trittmann
and was supported in part by the Department of Energy,
contract DE-FG02-98ER41087, and by grants of computing 
time from the Minnesota Supercomputing Institute.

\end{document}